\documentclass[%
 reprint,
superscriptaddress,
 amsmath,amssymb,
 aps,
]{revtex4-1}

\usepackage{graphicx}% Include figure files
\usepackage{dcolumn}% Align table columns on decimal point
\usepackage{bm}% bold math
\usepackage{booktabs}
\usepackage{longtable}

\def\hyph{-\penalty0\hskip0pt\relax}

\begin{document}

\preprint{APS/123-QED}

\title{New measurement of the $E_{\mathrm{c.m.}}=323$ keV resonance in the $^{19}$F$(p,\gamma)^{20}$Ne reaction}% Force line breaks with \\

\author{M. Williams}
 \email{mwilliams@triumf.ca}
\affiliation{Department of Physics, University of York, Heslington, York, YO10 5DD, United Kingdom}
\affiliation{TRIUMF, 4004 Wesbrook Mall, Vancouver, BC, Canada, V6T 2A3}
\author{P. Adsley}
\affiliation{School of Physics, University of the Witwatersrand, Johannesburg 2050, South Africa}
\affiliation{iThemba Laboratory for Accelerator Based Sciences, Somerset West 7129, South Africa}
%\author{U. Battino}
%\affiliation{University of Edinburgh, School of Physics and Astrophysics, Edinburgh EH9 3FD, UK}
%\affiliation{The NuGrid collaboration, http://www.nugridstars.org}
%\author{J. Jos\'e}
%\affiliation{Departament de F\'isica, Universitat Polit\`ecnica de Catalunya \& Institut d'Estudis Espacials de Catalunya (IEEC), C. Eduard Maristany 16, E-08019 \& Ed. Nexus-201, C. Gran Capit\`a, 2-4, E-08034, Barcelona, Spain}
\author{B. Davids}
\affiliation{TRIUMF, 4004 Wesbrook Mall, Vancouver, BC, Canada, V6T 2A3}
\affiliation{Department of Physics, Simon Fraser University, 8888 University Drive, Burnaby, BC, V5A 1S6, Canada}
\author{U. Greife}
\affiliation{Department of Physics, Colorado School of Mines, Golden, Colorado 80401, USA}
\author{D. Hutcheon}
\affiliation{TRIUMF, 4004 Wesbrook Mall, Vancouver, BC, Canada, V6T 2A3}
\author{J. Karpesky}
\affiliation{Department of Physics, Colorado School of Mines, Golden, Colorado 80401, USA}
\author{A. Lennarz}
\affiliation{TRIUMF, 4004 Wesbrook Mall, Vancouver, BC, Canada, V6T 2A3}
\author{M. Lovely}
\affiliation{Department of Physics, Colorado School of Mines, Golden, Colorado 80401, USA}
\author{C. Ruiz}
\affiliation{TRIUMF, 4004 Wesbrook Mall, Vancouver, BC, Canada, V6T 2A3}

\date{\today}% It is always \today, today,

\begin{abstract}

At temperatures below 0.1 GK the $^{19}$F$(p,\gamma)^{20}$Ne reaction is the only breakout path out of the CNO cycle. Experimental studies of this reaction are challenging from a technical perspective due to copious $\gamma$-ray background from the far stronger $^{19}$F$(p,\alpha)^{16}$O reaction channel. Here we present the first inverse kinematics study of the $^{19}$F$(p,\gamma)^{20}$Ne reaction, in which we measure the strength of the 323-keV resonance. We find a strength value of $\omega\gamma = 3.3^{+1.1}_{-0.9}$ meV, which is a factor of two larger than the most recent previous study. The discrepancy is likely the result of a direct to ground state transition which previous studies were not sensitive to. We also observe the transition to the first $2^{-}$ state, which has not been observed for this resonance in previous studies. A new thermonuclear reaction rate is calculated and compared with the literature. 

% The reaction is important for which environments.
% direct measurements of this reaction are made particularly difficult due to the dominant (p,ag) background.
% previous work only quote the strength to the first excited state.
% Here we use a recoil separator to identify the (p,g) reaction channel and show that much of the strength in fact goes to the third excited state or the ground state.
% We report a resonance strength of more than a factor of 2 larger than recent results.

\end{abstract}

\pacs{Valid PACS appear here}% PACS, the Physics and Astronomy
\maketitle

%\tableofcontents

\section{\label{sec:intro}Introduction}

%Read through Courture motivation, AGB stars etc. 
%Reach out to Jordi Jose on possible classical novae motivation.
% Perhaps also reach out to U. Battino
% Talk about previous work by courture

The CNO cycle is a catalytic process, in that 4 protons are converted to a helium nucleus without the loss of CNO seed material. In our own Sun the CNO cycle is only responsible for approximately 1\% of total energy production, with the remainder generated from the $pp$ chains \cite{adelberger2011}. For main sequence stars with masses greater than 1.5 $M_{\odot}$ the core temperature exceeds 20 MK and the CNO cycle becomes dominant over the $pp$ chains \cite{Rolfs-Book-1988}. At stellar temperatures, only the $^{19}$F$(p,\gamma)^{20}$Ne reaction can cause irreversible processing of CNO seed material towards the Ne-Na region. However, this reaction is in competition with the far stronger $^{19}$F$(p,\alpha)^{16}$O reaction. Therefore, the ratio of the $(p,\gamma)/(p,\alpha)$ reaction channels will largely determine the mass fraction of CNO material lost during hydrogen burning. Wiescher \textit{et al.} \cite{Wiescher1999} estimate that approximately 0.1\% of the $^{19}$F abundance produced at stellar temperatures is converted into Ne-Na material, which the authors show to be a non-negligible amount when integrated over the typical timescales for hydrostatic hydrogen burning. 

Studies of the $^{19}$F$(p,\gamma)^{20}$Ne reaction have proven experimentally challenging due to copious $\gamma$-ray background from $^{19}$F$(p,\alpha\gamma)^{16}$O. Earlier measurements have relied on detecting $E_{\gamma}>11$ MeV primary $\gamma$-rays to the first excited state in $^{20}$Ne ($E_{x} = 1.633$ MeV) \cite{keszthelyi1962,subotic1979}. These measurements were susceptible to pile-up from the 6.125-MeV $\gamma$-rays from $^{19}$F$(p,\alpha\gamma)^{16}$O. More recently, Couture \textit{et al.} employed a coincidence technique whereby 1.633-MeV $\gamma$-rays from the first excited state were measured by a HPGe clover detector in tandem with high energy $\gamma$-rays detected by NaI detectors \cite{couture2008}. A Q-value gating technique was then used to markedly reduce background from $^{19}$F$(p,\alpha\gamma)^{16}$O. However, none of the previous results were able to report any contribution from direct-to-ground state transitions. Here we report on a measurement performed in inverse kinematics, using the DRAGON recoil separator at TRIUMF. By selecting the reaction channel via tagging on the $^{20}$Ne recoils, we are able to measure the full reaction yield.

\section{\label{sec:exp}Experiment Description}

% Description of beam characteristics: species, charge state, energy, accelerators, intensity.

This study was performed using the Detector of Recoils And Gammas Of Nuclear reactions (DRAGON) \cite{Hutcheon-NIMPRSA498-2003}, located in the ISAC-I experimental hall \cite{laxdal2003} at TRIUMF, Canada's national laboratory for particle and nuclear physics. A sample of boron trifluoride was used to create the $^{19}\mathrm{F}$ beam utilizing the Multi Charge Ion Source (MCIS) \cite{Jayamanna2010}. The $q=4^{+}$ charge state was then extracted from the source and accelerated to an energy of $E_{lab} = 6.668$ MeV in the laboratory frame via the ISAC-I Radio-Frequency Quadrupole (RFQ) followed by the Drift-Tube Linac (DTL). The beam was delivered to the DRAGON experiment area with a typical intensity of $1 \times 10^{11}$ s$^{-1}$, and FWHM beam energy spread of $\Delta E/E \leqslant 0.4 \%$. 

A detailed description of the DRAGON facility, including its detectors and key components, is given in Refs. \cite{Hutcheon-NIMPRSA498-2003,williams2020} and so will not be repeated here. The only difference between the present set-up and that of Ref. \cite{williams2020} is that the second MCP (MCP1) was not part of the local time-of-flight system due to difficulties with excessive noise, which significantly lowered the detection efficiency. However, as described in Section \ref{sec:pid}, recoils could none-the-less be readily identified using signals from the first MCP (MCP0) and the DSSD.

\section{Data Analysis}

\subsection{\label{yield}Thick target yield and resonance strength}

The data analysis follows similarly to recently published work using the DRAGON facility aimed at determining absolute resonance strengths for narrow, isolated resonances \cite{williams2020}. The only difference in the analysis procedure in this work is the absence of any result obtained from heavy-ion only detection (`singles' events), due to a large quantity of scattered beam, from which genuine recoils could not be distinguished without a coincident $\gamma$-ray (`coincidence' events). The yield per incident beam ion is then given as:

%Laboratory experiments of nuclear reaction cross sections (and resonance strengths) measure the reaction \textit{yield}, which is defined per incident beam ion as:

%\begin{equation} \label{eqn:yield1}
%	Y = \frac{N^{tot}_{r}}{N_{b}}
%\end{equation}

%where $N^{tot}_{r}$ is the total number of reactions that occur, and $N_{b}$ is the number of beam ions incident on the target. At DRAGON, the total number of reactions is inferred by combining the number of detected recoils with the systematics of the experiment. Therefore, Equation \ref{eqn:yield1} can be re-written as:

\begin{equation} \label{eqn:yield2}
	Y = \frac{N^{det}_{r}}{N_{b} \, \varepsilon_{\mathrm{coinc}}}
\end{equation}

where $\varepsilon_{\mathrm{coinc}}$ is the product of all efficiencies affecting the number of detected coincident recoil-$\gamma$ events, $N^{\mathrm{det}}_{r}$. The total coincidence detection efficiency pertaining is given as:

%\begin{equation} \label{eqn:singles_eff}
%\varepsilon_{\textrm{DRA}}^{\mathrm{sing}} = f_{q} \cdot \tau_{\mathrm{MCP}} \cdot \varepsilon_{\mathrm{MCP}} \cdot \varepsilon_{\mathrm{DSSD}} \cdot \tau_{\mathrm{rec}} \cdot \lambda_{\mathrm{tail}}
%\end{equation}

\begin{equation} \label{eqn:coinc_eff}
\varepsilon_{\mathrm{coinc}} = f_{q} \cdot \tau_{\mathrm{MCP}} \cdot \varepsilon_{\mathrm{MCP}} \cdot \varepsilon_{\mathrm{DSSD}} \cdot \varepsilon_{\gamma} \cdot \lambda_{\mathrm{coinc}}
\end{equation}

The first four terms in Equation \ref{eqn:coinc_eff} are not particular to coincidence events and would also apply equally to the detection efficiency of heavy-ion only events. These are: the recoil charge state fraction $(f_{q})$, MCP transmission efficiency $(\tau_{\mathrm{MCP}})$, MCP detection efficiency $(\epsilon_{MCP})$ and detection efficiency of the DSSD $(\varepsilon_{\mathrm{DSSD}})$. $\lambda_{\mathrm{coinc}}$ is the live time fraction where both the `head' DAQ, responsible for generating triggers from the BGO array, and focal plane detector (`tail') DAQ are able to accept new triggers \cite{Greg2014}. The recoil-gamma coincidence efficiency $(\varepsilon_{\gamma})$ is the probability that a transmitted recoil will be recorded in coincidence with a prompt $\gamma$-ray detected by the BGO array. This quantity is obtained via simulation, calculated as:

%The recoil transmission efficiency, $\tau_{\mathrm{rec}}$, relates to the number of recoils that are produced within the acceptances of the separator. Obtained through simulation, this quantity depends on the kinematics of the radiative capture reaction and its effect on the transmission of recoils through the separator. The recoil-gamma coincidence efficiency $(\varepsilon_{\gamma})$ is the probability that a transmitted recoil will be recorded in coincidence with a prompt $\gamma$-ray detected by the BGO array. This quantity is also obtained via simulation, calculated as:

\begin{equation} \label{eqn:sim_eff1}
\varepsilon_{\gamma} = \frac{N^{\mathrm{sim}}_{\mathrm{coinc}}}{N^{\mathrm{sim}}_{\mathrm{react}}},
\end{equation}

where $N^{\mathrm{sim}}_{\mathrm{react}}$ is the simulated number of reactions, and $N^{\mathrm{sim}}_{\mathrm{coinc}}$ is the total number of $\gamma$-rays detected in coincidence with a recoil transmitted to the focal plane. The transmission of heavy-ions to the focal plane, $\tau_{r}$, without necessarily being accompanied with a detected $\gamma$-ray, can also be obtained through simulation as:  

\begin{equation}
\tau_{r} = \frac{N^{\mathrm{sim}}_{\mathrm{sing}}}{N^{\mathrm{sim}}_{\mathrm{react}}},
\end{equation}

where $N^{\mathrm{sim}}_{\mathrm{sing}}$ is the number of detected heavy-ion events at the focal plane in the simulation. Note that the definition of the recoil-$\gamma$ coincidence efficiency in Equation \ref{eqn:sim_eff1} already accounts for the transmission of recoils, therefore, $\tau_{\mathrm{rec}}$ need not be included in the total coincidence efficiency.

For narrow resonances, wherein the resonance width is small compared to the target width, the reaction yield becomes the thick target yield $(Y \rightarrow Y_{\infty})$. The $E_{\mathrm{c.m.}}=323$-keV resonance is reported as having a total width of 2 keV \cite{couture2008}, which is sufficiently smaller than the center-of-mass target thickness here of $\approx 24$ keV to satisfy the thick target assumption. For a narrow Breit-Wigner resonance the thick target yield is related to the resonance strength by:

\begin{equation} \label{eqn:wg}
\omega \gamma = \frac{2Y_{\infty}}{\lambda_{r}^{2}} \frac{m_{t}}{m_{t}+m_{b}} \, \epsilon_{\mathrm{lab}},
\end{equation}

where $\omega\gamma$ is the resonance strength in eV, $m_{t}$ and $m_{b}$ are the target (proton) and beam ($^{19}$F) masses (in $u$) respectively, $\epsilon_{\mathrm{lab}}$ is the laboratory frame stopping power (eV/cm$^{2}$), and $\lambda_{r}$ is the de Broglie wavelength (cm) associated with the relative energy of the resonance in the center of mass frame.

% include figure showing stopping power measurements as a function of beam energy, with a comparison to SRIM. Devin shows the same thing in his paper.

%Furthermore, the results presented here are largely independent from precise knowledge of the $\gamma$-decay cascade from the compound nuclear resonance under investigation. The same is also true for non-resonant yield measurements, in that the identification all direct capture transitions need not be precisely measured, since the number of detected recoils of interest is representative of the entire yield across the target.

\subsection{\label{sec:beam_energy}Beam energy and stopping power}

The incident beam energy was measured by tuning through the first magnetic dipole (MD1) onto a downstream pair of slits. The energy of the beam, $E$, tuned to pass through the first dipole magnet (MD1) is related to the MD1 field, as measured by its respective NMR probe, with the following equation taken from Ref. \cite{Hutcheon-NIMPRA-2012}: 

%The slit plates are electrically isolated so as to enable current to be measured on each plate. The slit plates are unsuppressed however, and therefore do not permit measurement of absolute current. Nonetheless, with the slits closed to 2 mm, they do serve as accurate beam tuning diagnostics to center a given charge state through MD1.

\begin{equation} \label{eqn:energy}
E_{b}/m_{b} = c_{\textrm{mag}} (qB/m_{b})^{2} - \frac{1}{2uc^{2}}(E/m_{b})^{2},
\end{equation}

where $m_{b}$ is the atomic mass of the beam in $u$, $q$ is the beam charge state after the target, $B$ is the MD1 field (in Tesla) measured by its NMR probe, $u$ is the atomic mass unit, $c$ is the the speed of light, and $c_{\textrm{mag}}=48.15 \pm 0.07 \; \textrm{MeV} \; \textrm{T}^{2}$ is a constant related to the effective bending radius of MD1 \cite{Hutcheon-NIMPRA-2012}. The final term is a relativistic correction that can be neglected. 

The total energy loss across the gas target was measured by using Equation \ref{eqn:energy} to determine the beam energy with and without gas present in the target. The stopping power across the target can be directly obtained by combining the measured energy loss and target number density. Here we measure a beam stopping power of $\epsilon_{\mathrm{lab}} = 86.5 \pm 3.7$ eV/($10^{15}$ atoms / cm$^{2}$). This value is in almost exact agreement with the prediction from SRIM-2013 of 86.4 eV/($10^{15}$ atoms / cm$^{2}$) \cite{ZIEGLER2010}.

\subsection{\label{sec:beamnorm}Beam Normalization}

The beam normalization procedure follows similarly to that described in Ref. \cite{williams2020}. The core elements will none-the-less be summarized here for convenience. Two silicon surface barrier (SSB) detectors mounted inside the gas target at $30^{\circ}$ and $57^{\circ}$ relative to the beam axis measure the scattering rate of target ions. For a fixed beam energy, the scattering rate is proportional to the beam current and the target gas pressure, the former is measured by hourly Faraday cup readings. For each cup reading, a normalisation coefficient is calculated, which relates the SSB rates to the beam current and target pressure:   

%The total number of incident beam ions was determined by taking hourly beam current measurements using a Faraday cup (FC4) positioned approximately 2 m upstream of the gas target. Beam fluctuations within each data taking run were accounted for by relating these regular current measurements to the number target atoms scattered into two ion implanted silicon (IIS) detectors, mounted at $30^{\circ}$ and $57^{\circ}$ relative to the beam axis. The beam normalization coefficient, R, for a given run, is obtained as:

\begin{equation} \label{eqn:rfactor}
R = \frac{I}{e q} \frac{\Delta t \, P}{N_{p}} \, \varepsilon_{t}
\end{equation}

where $I$ is the beam current as measured by FC4 (situated upstream of the target) and $e  q$ is the charge of the incident beam ions. $\Delta t$ is a short time interval, immediately proceeding a Faraday cup reading, over which the target pressure $P$ and number of elastically scattered protons $N_{p}$ is measured. The beam transmission efficiency $(\varepsilon_{t})$ through the target apertures is measured, prior to filling the target with hydrogen, by recording the ratio of current measured by FC1 (immediately downstream of the target) over the current measured by FC4. The average normalization coefficient over all runs, $\langle R\rangle$, can then be used with Equation \ref{eqn:nbeam} to determine the total number of beam ions:

\begin{equation} \label{eqn:nbeam}
N_{b} = \frac{\langle R\rangle N_{p}^{\textrm{tot}}}{\langle P \rangle},
\end{equation}

where $N_{p}^{\textrm{tot}}$ is now the total number of elastically scattered protons, and $\langle P \rangle$ is the average pressure measured over all runs. Using the method explained in this section, we find a total number of $N_{b} = (7.035 \pm 0.186) \times 10^{15}$ beam ions on target over the course of the $E_{\mathrm{c.m.}}=323$-keV yield measurement.

\subsection{\label{sec:csd} $^{20}\textrm{Ne}$ Charge State Fraction}

\begin{figure}[h!]
 \includegraphics[width=0.45\textwidth]{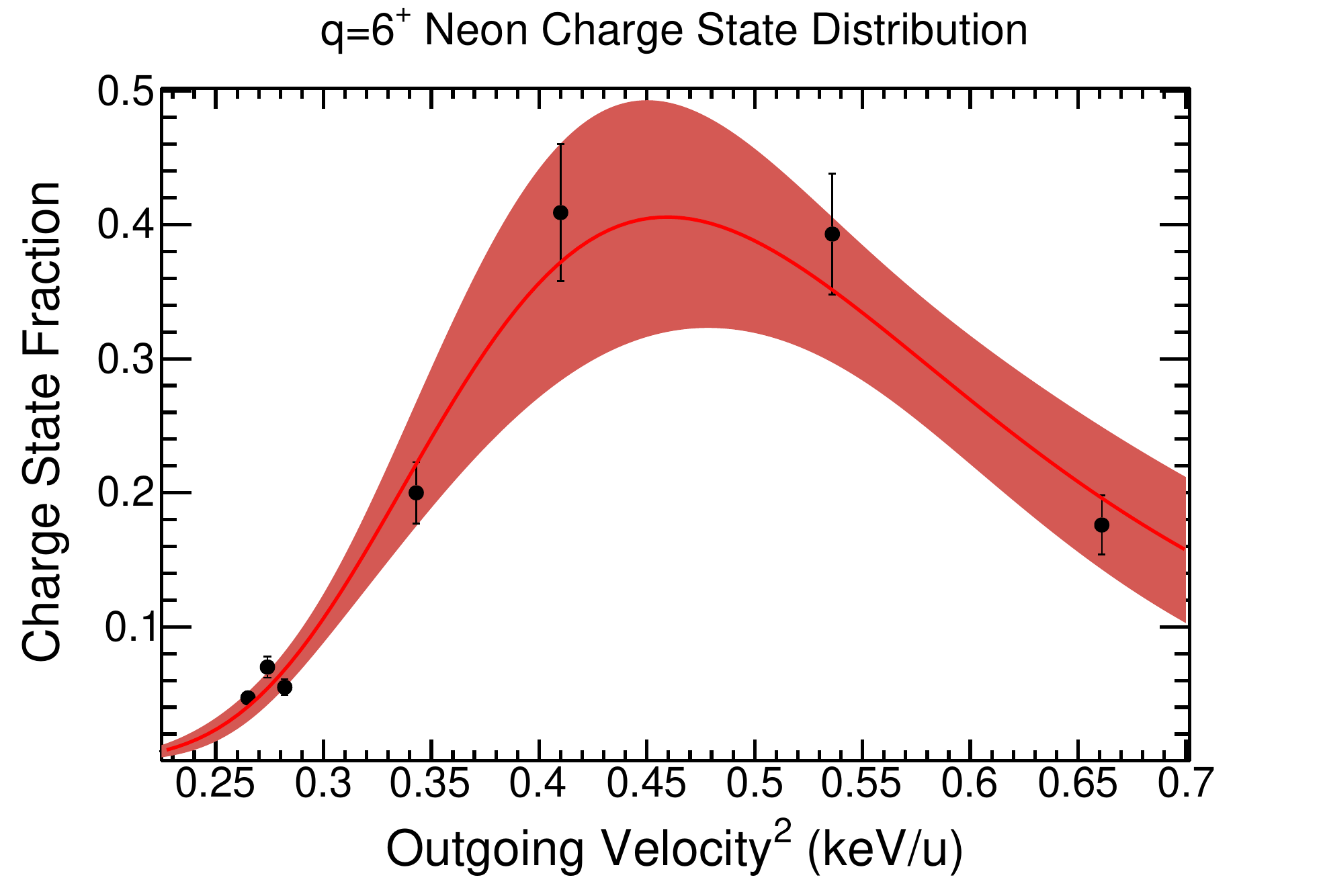}
\caption{Normalized q=6+ charge state fraction as a function of outgoing $^{20}$Ne velocity$^{2}$. The distribution is fit with the semi-empirical formula of Liu \textit{et al.} \cite{wenji}, with the shaded region indicating the $1\sigma$ confidence limits of the fit.} \label{fig:csd}
 \end{figure}

DRAGON is designed to accept only a single charge state through the separator to the focal plane detectors. Therefore, in order to recover the full reaction yield, the charge state fraction of the $q=6^{+}$ $^{20}$Ne recoils to which DRAGON is tuned to accept must be known. For the present work, data from previously measured charge state distributions using neon beams at DRAGON were taken from Ref. \cite{lovely_2020}. The data were fit with Gaussian distributions normalized to unity and the resulting $q=6^{+}$ charge state fraction as a function of the mean outgoing beam velocity$^{2}$ was fit with the semi-empirical formulae of Lui \textit{et al.} \cite{wenji}. The $q=6^{+}$ charge state fraction as a function of outgoing beam velocity$^{2}$ is shown on Figure \ref{fig:csd}. For the measurement of the 323-keV resonance the outgoing $^{20}$Ne recoil velocity$^{2}$ was 292 keV/$u$, which from the fit shown in Figure \ref{fig:csd} implies a 6+ charge state fraction of $\varepsilon_{\mathrm{CSD}} = (8.9 \pm 1.6)\%$. Initially, DRAGON was set to accept the more abundantly produced $q=5^{+}$ recoil charge state, but the scattered beam rate (`leaky-beam') was deemed too high for the DSSD to sustain over the experiment. Therefore, to lower the scattered beam rate, the tuned charge state was increased to $6^{+}$ at the expense of a reduced overall efficiency.

\subsection{Heavy-ion Detection Efficiency}

For this work the heavy-ion detectors located at the focal plane consisted of an MCP followed by a DSSD. The MCP detects secondary electrons emitted as the ions traverse a thin diamond-like carbon foil inserted into the beam line. The electrons are deflected towards the MCP by electric fields generated by set of wire grids held at bias. The foils and wire grids result in some losses in transmission to the DSSD, which can be determined using successive attenuated beam runs measuring the DSSD rate with the MCP inserted and retracted from the beam line. The MCP transmission is measured as $\tau_{\mathrm{MCP}} = (90.0 \pm 1.3)\%$. The MCP detection efficiency, also measured during attenuated beam runs, was essentially perfect at 99.99\%. The DSSD has a geometric efficiency of $\varepsilon = (96.15 \pm 0.53) \%$ \cite{Wrede-NIMB204-2003}.

\subsection{Particle ID} \label{sec:pid}

\begin{figure}[h!]
\centering
\minipage{0.5\textwidth}
 \includegraphics[width=1.\textwidth]{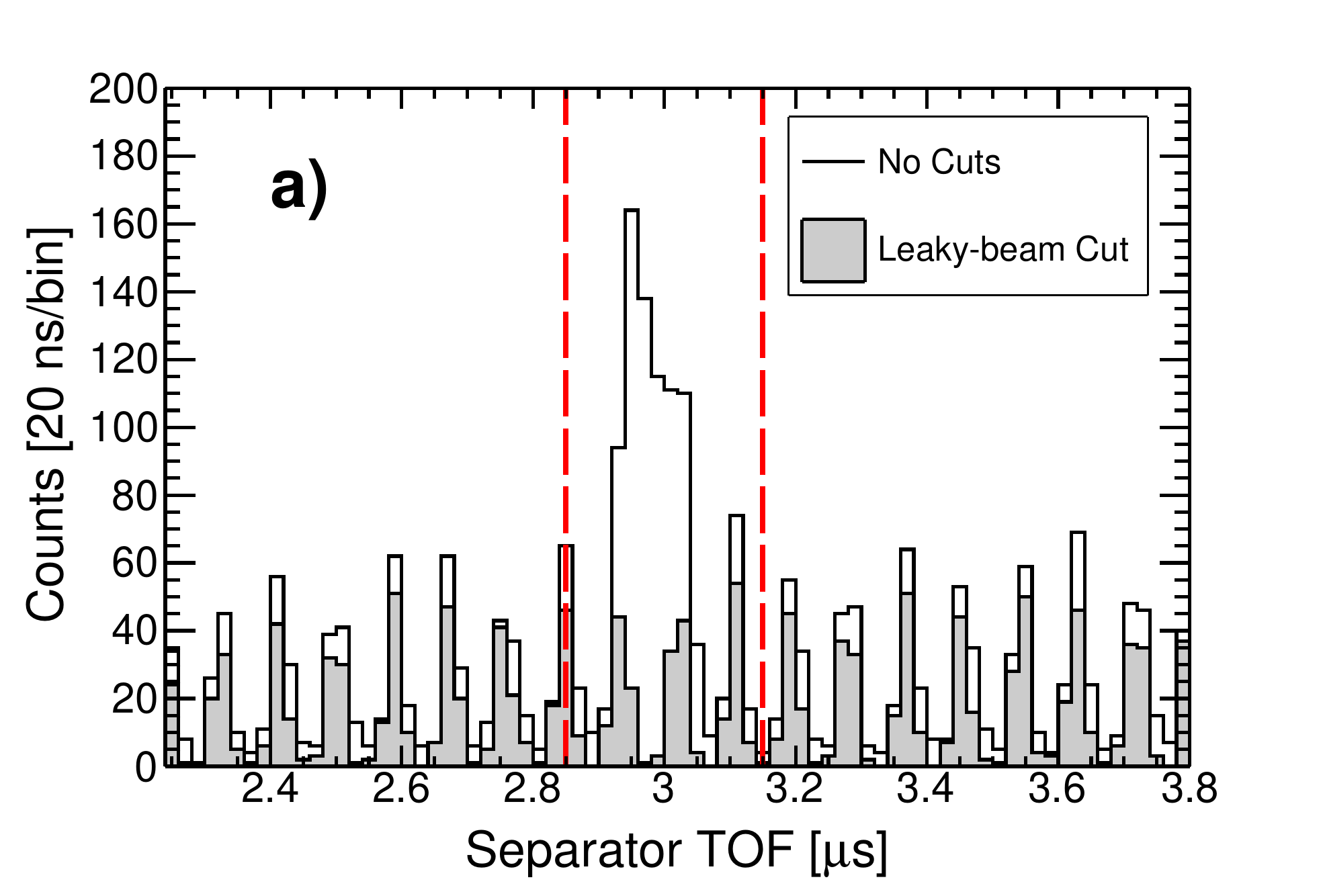} \label{fig:septof_cuts}
 \endminipage\hfill
\minipage{0.5\textwidth}
 \includegraphics[width=1.\textwidth]{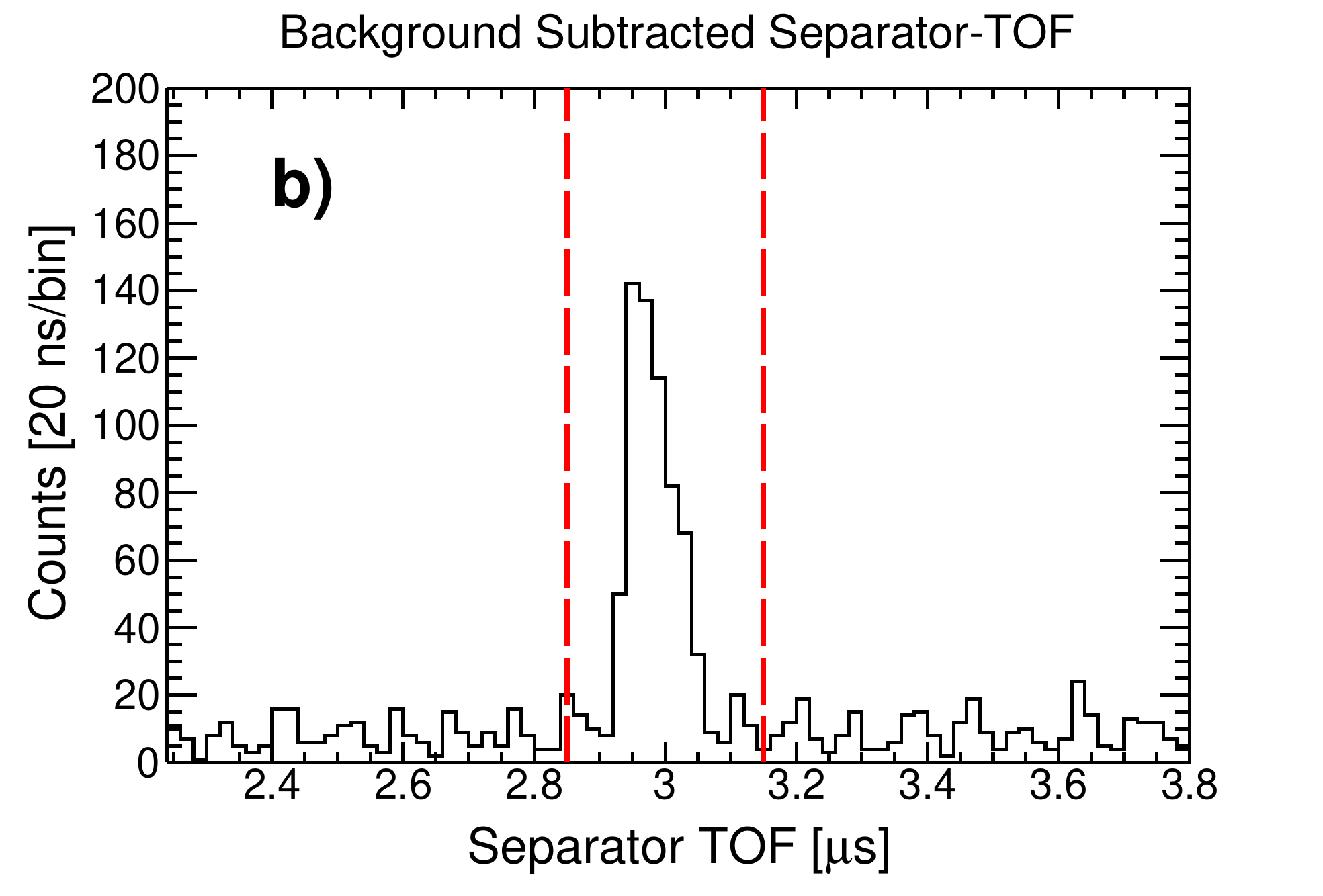} \label{fig:septof_bgsub}
 \endminipage\hfill
\caption{a) Separator time-of-flight with no heavy-ion gates and gated on the leaky beam locus (gray shaded histogram) displayed in Figure \ref{fig:mcpVdssd}. b) Background-subtracted separator-TOF. The separator-TOF signal region is bounded by the red vertical dashed lines. } \label{fig:septof}
 \end{figure}

The primary method for identifying genuine recoil-$\gamma$ events is to inspect the separator time-of-flight (TOF) spectrum, that is, the time difference between BGO and heavy-ion events. The raw separator-TOF is shown in Figure \ref{fig:septof}. There is a prominent peak at around 3 $\mu$s, surrounded by a comb-like pattern of peaks repeated every 86 ns. These smaller peaks are explained as leaky beam events, and their periodicity is the result of the pulsed time structure of the delivered beam.

\begin{figure}[h!]
\centering
\minipage{0.5\textwidth}
 \includegraphics[width=1.\textwidth]{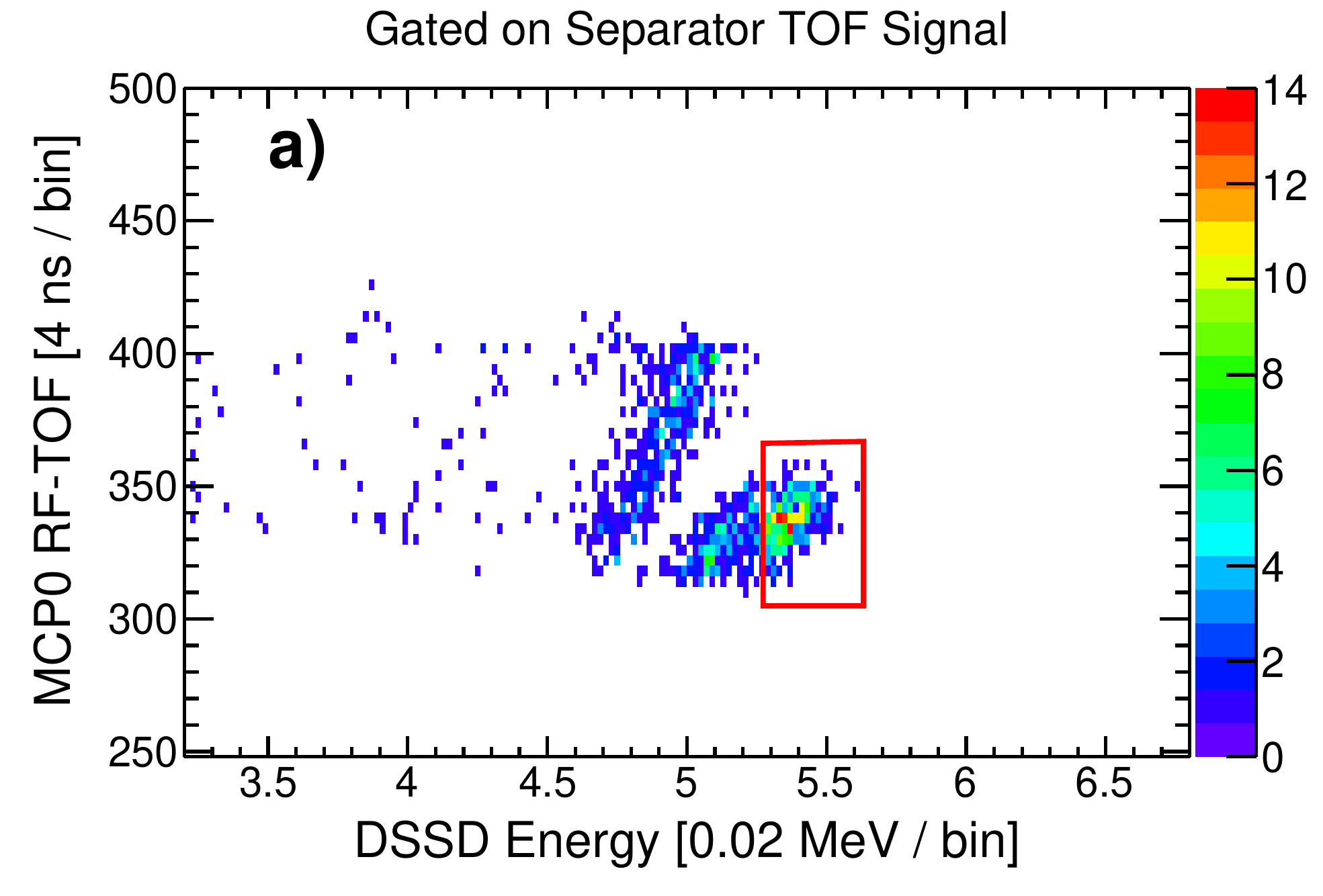} \label{fig:mcpVdssd_septof}
\endminipage\hfill
\minipage{0.5\textwidth}
 \includegraphics[width=1.\textwidth]{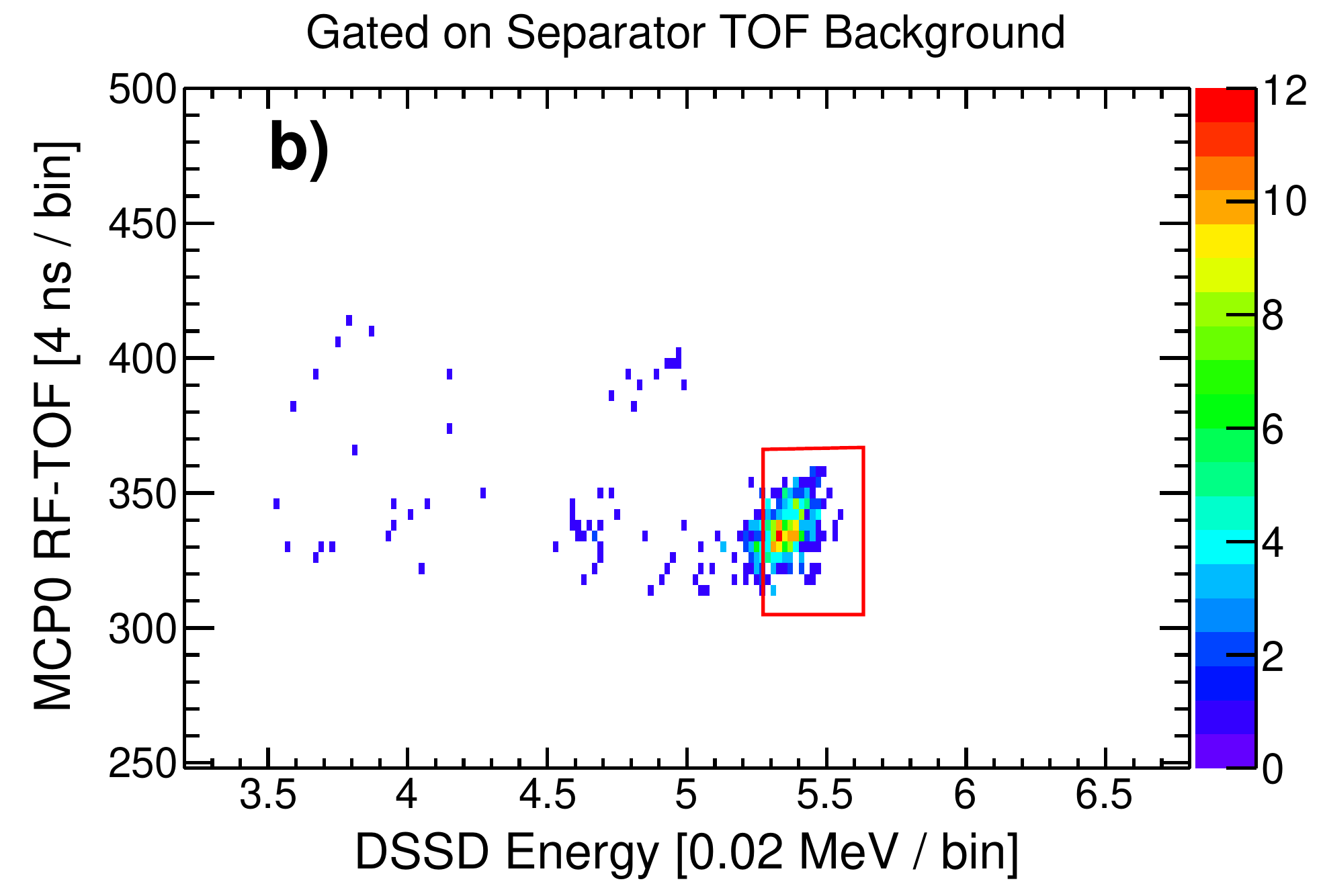} \label{fig:mcpVdssd_bckgrnd}
\endminipage\hfill
\caption{MCP RF-TOF plotted against DSSD energy with a) a gate around the separator TOF signal region indicated in Figure \ref{fig:septof}, and b) a gate on a background region of the separator TOF spectrum between 3.2 and 3.6 $\mu$s. The red rectangle indicates a cut around the leaky beam events.}
 \label{fig:mcpVdssd}
\end{figure}

The background can be significantly reduced by gating on the MCP-RF time-of-flight vs DSSD energy. The MCP-RF TOF is the time between an MCP event and the leading-edge discriminated RF signal. Figure \ref{fig:mcpVdssd}a shows MCP RF-TOF vs DSSD energy gated on the separator-TOF signal region in Figure \ref{fig:septof} and Figure \ref{fig:mcpVdssd}b shows the same but gated on a similar size region of the separator-TOF background. It is clear that the majority of the leaky-beam is concentrated in a single locus. The recoils however appear to have a larger time spread, large enough in-fact that there appears to be some wrap-around effect in the MCP RF-TOF as events become associated with the next pulse of the discriminated RF signal. This RF wrap-around effect makes it difficult to place a cut around the region of interest. However, the background in the separator-TOF spectrum can be markedly reduced by excluding events occurring within the leaky-beam locus. The final number of recoils is found by subtracting the average background from the number of events contained in the signal region between 2.85 and 3.15 $\mu$s. The average background was estimated by sampling the uniform background outside of the signal region. The final number of recoils is found to be $514^{+27}_{-26}$. Confidence bounds were evaluated at $1\sigma$ using the Rolke method \cite{rolke2001}, assuming a Poisson background model. 

\subsection{Coincidence Efficiency}

The present work suffered from too much leaky-beam background at the focal plane to extract a result from detection of heavy-ions alone; coincident $\gamma$-rays are required to identify the recoils of interest. Therefore, the efficiency of the BGO array must be accounted for in addition to the fraction of recoils that make it through the separator. Taken together, these give the coincidence efficiency of Equation \ref{eqn:coinc_eff}, and is obtained via simulation. A \texttt{GEANT3} simulation of the dragon facility is used to model both the interaction of $\gamma$-rays with the BGO array as well as the transport of recoils through the separator, taking into account the reaction kinematics. A software threshold of 2 MeV is imposed on the data, therefore the simulation should acquire the fraction of reactions that give rise to a detected recoil in coincidence with at least one $\gamma$ ray above 2 MeV.

\begin{figure}[h!]
 \includegraphics[width=0.5\textwidth]{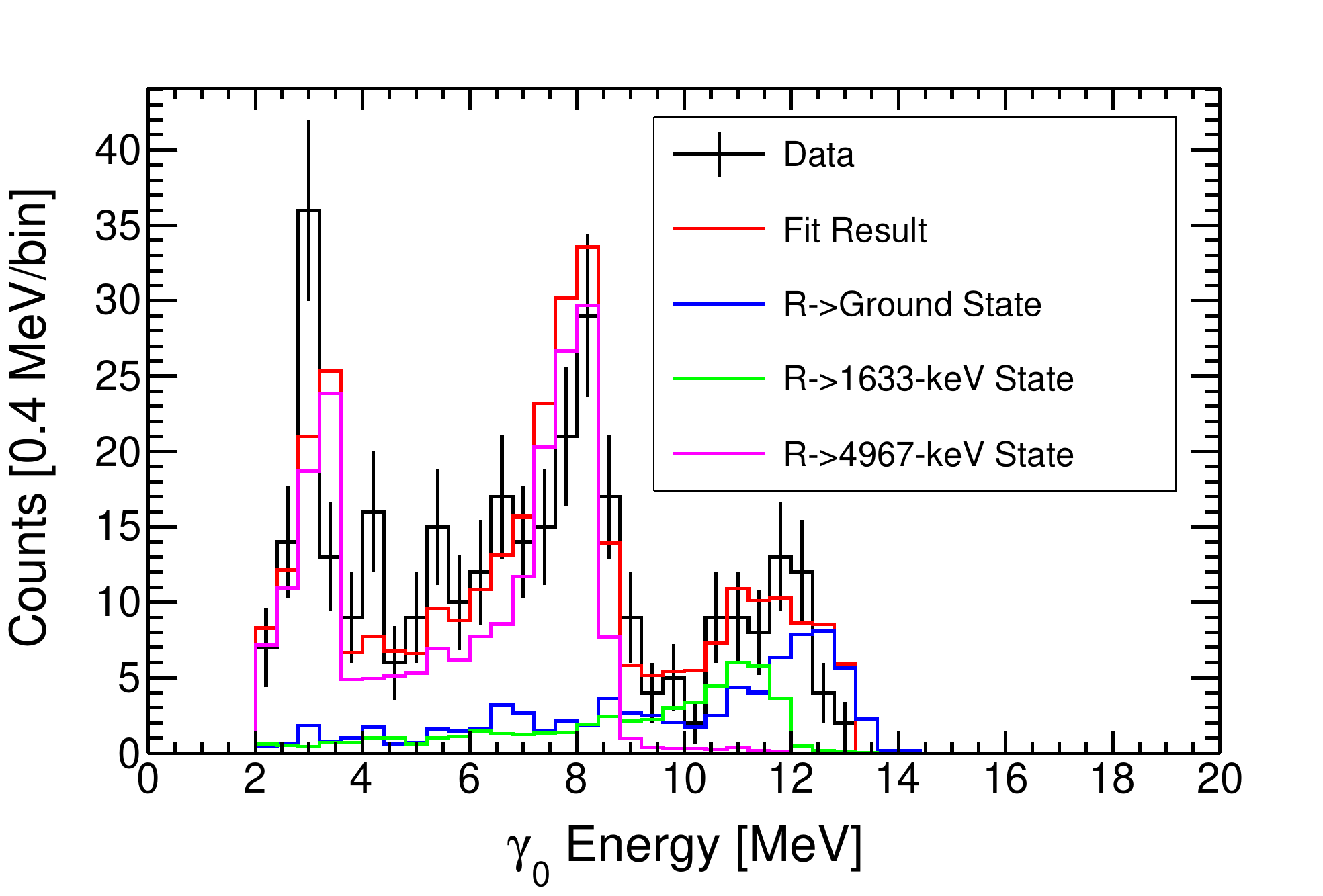}
\caption{BGO spectra for $\gamma$-rays detected in coincidence with $^{20}$Ne recoils. The data is fit with simulated spectra for each primary decay branch. The spectrum is dominated by decays through the 4966.5-keV state. At high energies there appears to be two bumps arising from decay to the first excited state and then to the ground state.} \label{fig:bgo}
 \end{figure}

There are no published branching ratios for this resonance. Instead, the branching ratios were estimated by simulating possible decay branches and fitting with the observed BGO spectra using the \texttt{TFractionFitter} class included in the \texttt{ROOT} analysis package \cite{brun1997root}. From the work of Couture \textit{et al.} \cite{couture2008}, it is clear that the transition from the first excited state to the ground state is populated for this resonance. For the 634-keV resonance, Couture \textit{et al.} also observed the 3.33-MeV $\gamma$-ray from the $2^{-}$ state at $E_{x}=4966.5$ keV. However, no yield for this $\gamma$-ray was reported for the 323-keV resonance and was therefore neglected. In contrast, this work reveals a clear branch to the $E_{x}=4966.5$-keV state as demonstrated by Figure \ref{fig:bgo}. It is worth noting here that the 4966.5-keV state decays predominantly to the first excited state, therefore the resonance strength reported by Couture \textit{et al.} would largely subsume the strength from this additional decay and not be significantly impacted by observation of this branch. This work also suggests a contribution from a direct-to-ground state branch, which no other study of the $^{19}$F$(p,\gamma)^{20}$Ne reaction would be sensitive to. The fractions calculated by the \texttt{TFractionFitter} need to be corrected by the differing recoil-$\gamma$ coincidence efficiencies for each decay branch to recover the true branching ratios. The \texttt{TFractionFitter} result, recoil-$\gamma$ coincidence efficiency, and true branching ratio for each primary decay are given in Table \ref{tab:branches}. A simulation of the final branching ratios given in Table \ref{tab:branches} gives a coincidence efficiency of $\varepsilon_{\gamma} = 24^{+5}_{-7}$ \%. The systematic uncertainty in the GEANT3 simulation is taken to be $10\%$, as determined with various calibrated $\gamma$-ray sources \cite{bgo_sim}. However, the uncertainty in this case is larger to account for the uncertainty in the $\gamma$-decay branching ratios, which each result in very different recoil transport efficiencies through DRAGON. 

\begin{table}[h!]
    \centering
    \begin{tabular}{l c c c}
    \toprule[1pt]\midrule[0.3pt]
	\noalign{\smallskip}
    Branch  &  Fit Result (\%) &  $\varepsilon_{\gamma}$ (\%) & BR (\%) \\ \hline
            \noalign{\smallskip}
    G.S.     & 23.1(6.1) & 9.6(1.0)  & 57.5(18.1) \\
    1633-keV & 14.8(5.5) & 21.0(2.2) & 16.9(6.9)  \\
    4967-keV & 62.1(5.4) & 58.1(6.0) & 25.6(4.8)  \\
    \noalign{\smallskip} 
	\midrule[0.3pt]\bottomrule[1pt]
    \end{tabular}
    \caption{Observed branching ratios obtained by fitting contributions of each branch to the data. The actual branching ratios are obtained by accounting for the differing recoil-$\gamma$ coincidence efficiencies for each branch.}
    \label{tab:branches}
\end{table}

The possibility of a large ground state branch was discounted by Keszthelyi \textit{et al.} \cite{keszthelyi1962}, as the authors assumed that this branch would be similarly suppressed as with the 634-keV state, which also has a $1^{+}$ spin-parity assignment. However, the same authors remark that a comparison of the exit channel partial widths suggests a very different structure for the two resonances. The 323-keV resonance is characterized by a very large alpha width and small proton width, whereas the 634-keV resonance has the opposite behaviour. Keszthelyi \textit{et al.} give a tentative explanation for this by suggesting that the 13.196 MeV state might be the result of a $^{16}$O + $\alpha$ cluster state and the 13.511 MeV level that of a $^{19}$F + $p$ state. Given the potentially different structure of the two states it is not clear why the absence of a measured ground state $\gamma$-decay branch for the 634-keV resonance would necessarily allow one to assume the same for the 323-keV resonance. 

We note that M1 transition probabilities in $^{20}$Ne were studied by Bendel \textit{et al.} \cite{bendel1971} via inelastic electron scattering. The $(e,e^{\prime})$ reaction measured at $180^{\circ}$ is highly selective for M1 excitation. In even-even $T=0$ nuclei, states with $T=1$ and $J^{\pi}=1^{+}$, such as the resonance studied in this work, will be preferentially excited. As Bendel \textit{et al.} note, no signal was observed for the states at 13.196 or 13.511-MeV, the former giving rise to the 323-keV resonance and which are both $T=1$ and $J^{\pi}=1^{+}$. However, the total radiative-widths of $^{20}$Ne states above the proton threshold are relatively small. The present work suggests a ground state radiative-width that is approximately equivalent to $\gamma$-decay resulting in de-excitation of the first $2^{+}$ state (which includes transitions to the first $2^{-}$ state), reported by Couture \textit{et al.} as 0.1 eV \cite{couture2008}. This radiative width is a factor of 100 smaller than the strong $T=1$ $J^{\pi}=1^{+}$ state at 11.2 MeV, which dominates the spectra shown in Figures 1 and 2 of Bendel \textit{et al.} with $\Gamma_{\gamma_{0}} = 11.2^{+2.1}_{-1.8}$ eV \cite{bendel1971}. Since the differential cross section of inelastic electron scattering scales with the reduced M1 transition probability, and hence the radiative width, then for transitions between states of the same spin one would expect a signal amplitude of $\approx2\%$ that of the 11.2 MeV state (since the $\gamma$-ray energy is a factor 1.2 greater and the M1 transition probability goes as $E_{\gamma}^{3}$). Such a small signal would be readily obscured by background present in the spectra shown by Bendel \textit{et al} \cite{bendel1971}, which is relatively high and increases toward higher excitation energies.

Another consideration for determining the coincidence efficiency is the DAQ livetime. The event rate in the BGO array was dominated throughout the experiment by the far stronger $^{19}$F$(p,\alpha\gamma)^{20}$Ne reaction. This had the effect of significantly lowering the DAQ livetime. Typically, the coincidence livetime fraction for DRAGON experiments is in the range of 80 - 85\%, for this experiment however, the coincidence livetime fraction was calculated as $\lambda_{\mathrm{coinc}} = 39.504 \pm 0.002$ \%.

\section{\label{sec:results}Results}

The final strength value for the 323-keV resonance is determined using Equation \ref{eqn:wg} and the values listed in Table \ref{tab:results}, from which we obtain $\omega\gamma = 3.3 ^{+1.1}_{-0.9}$ meV. This is more than a factor of two stronger than the $1.4 \pm 0.4$ meV strength reported by Couture \textit{et al.} \cite{couture2008}, albeit within $2\sigma$ error. Couture \textit{et al.} assumed that most of the $\gamma$-decay proceeds via a cascade of $\gamma$-rays passing through the first excited state in $^{20}$Ne. However, this work suggests a significant branch to the ground state, which Couture \textit{et al.} would not have been sensitive to, due to the requirement of a coincidence with the 1.633-MeV $\gamma$-ray. Including the unseen $(58 \pm 18)$\% contribution from a direct to ground state branch would give a total strength of $3.3$ meV for that work.

\begin{table}[h!]
    \centering
    \begin{tabular}{l  r}
    \toprule[1pt]\midrule[0.3pt]
	\noalign{\smallskip}
    Number of beam ions, $N_{b}$     &  $(7.035 \pm 0.186) \times 10^{15}$ \\
    Number of recoils, $N_{r}$ & $514^{+27}_{-26}$ \\
    Stopping Power, $\epsilon_{\mathrm{lab}}$ &  $(86.5 \pm 3.7)$ eV/($10^{15}$ atoms/ cm$^{2}$) \\
    de Broglie Wavelength, $\lambda^{2}_{r}$ &  $2.663 \times 10^{-23}$ cm$^{2}$ \\
    Charge state fraction, $f_{q}$ &  $(8.9 \pm 1.6)$ \% \\
    Recoil-$\gamma$ Efficiency, $\varepsilon_{\gamma}$ &  $24^{+5}_{-7}$ \% \\
    DSSD Efficiency, $\varepsilon_{\mathrm{DSSD}}$ & $(96.15 \pm 0.53)$ \% \\
    MCP Transmission, $\tau_{\mathrm{MCP}}$ &  $(90.0 \pm 1.3) \%$ \\
    DAQ Livetime, $\lambda_{\mathrm{coinc}}$ &  $(39.504 \pm 0.002)$ \%\\
    \noalign{\smallskip} 
	\midrule[0.3pt]\bottomrule[1pt]
    \end{tabular}
    \caption{Values used to the determine the 323-keV resonance strength.}
    \label{tab:results}
\end{table}

Previous measurements of this resonance have been performed by Suboti\'c \textit{et al.} \cite{subotic1979} and Keszthelyi \textit{et al.} \cite{keszthelyi1962}. The study by Keszthelyi \textit{et al.} presented data from two separate techniques to measure the ratio of high energy $\gamma$-rays from $^{19}$F$(p,\gamma)^{20}$Ne to 6.1 MeV $\gamma$-rays from $^{19}$F$(p,\alpha\gamma)^{16}$O. The first method involved counting $\gamma$ rays in the two energy regions of interest, but was strongly hampered by pile-up of 6.1-MeV $\gamma$ rays obscuring the 11.6-MeV $\gamma$ rays from the resonance decay to the first excited state in $^{20}$Ne. The second measurement, which dominates the weighted average used to arrive at the adopted result, utilized threshold activation of the $^{121}$Sb$(\gamma,n)^{120}$Sb reaction ($^{121}$Sb neutron separation energy, $S_{n} = 9.252$ MeV). However, this latter measurement would not be able to determine the relative contribution of the direct-to-ground state vs decay to the first excited state, and neither method would be able to detect any decay strength to the 4966.5 keV state. A resulting ratio in the partial widths of $\Gamma_{\gamma}/\Gamma_{\alpha_{2}} = 1 \times 10^{-4}$ was obtained. The $\alpha_{2}$-width from the 323-keV state was previously reported as 2800 eV, but was revised down to $1971^{+369}_{-294}$ keV by Couture \textit{et al.} \cite{couture2008}. Applying the aforementioned partial width ratio to the widths reported in Couture \textit{et al.} gives a resonance strength of $2.5 \pm 0.9$ meV. Correcting for the missing 25.6\% branch to the 4.9665 MeV state would then give a full resonance strength of $3.4 \pm 1.2$ meV, which is well within error of the present work.

The measurement by  Suboti\'c \textit{et al.} \cite{subotic1979} relied on measuring the 323-keV resonance yield relative to that of the 634-keV resonance, assuming a strength value for the latter of 1.61 eV. However, there is considerable disagreement in the value of $\Gamma_{\gamma}$ for the 634-keV resonance, particularly when comparing the work of Couture \textit{et al.} \cite{couture2008} to previous literature values. Adopting the widths of Couture \textit{et al.} would revise the strength of the 634-keV resonance down to 0.93 eV. This in turn would lower the strength for the 323-keV resonance reported by Suboti\'c \textit{et al.} down from 10 meV to 5.8 meV. However, factoring in the unseen direct to ground state transition would revise the strength back up to 13.8 meV. Further inspection of Table 1 in Suboti\'c \textit{et al.} \cite{subotic1979} reveals that work to be highly discrepant compared with both previous and subsequent literature values for every resonance investigated. The cause for this discrepancy may be due to pile-up from $^{19}$F$(p,\alpha\gamma)^{16}$O. We therefore caution against using the work of Suboti\'c \textit{et al.} in compiling resonance parameters for the $^{19}$F$(p,\gamma)^{20}$Ne reaction.

\section{\label{sec:rate}Thermonuclear Reaction Rate}

\begin{figure}[h!]
 \includegraphics[width=0.5\textwidth]{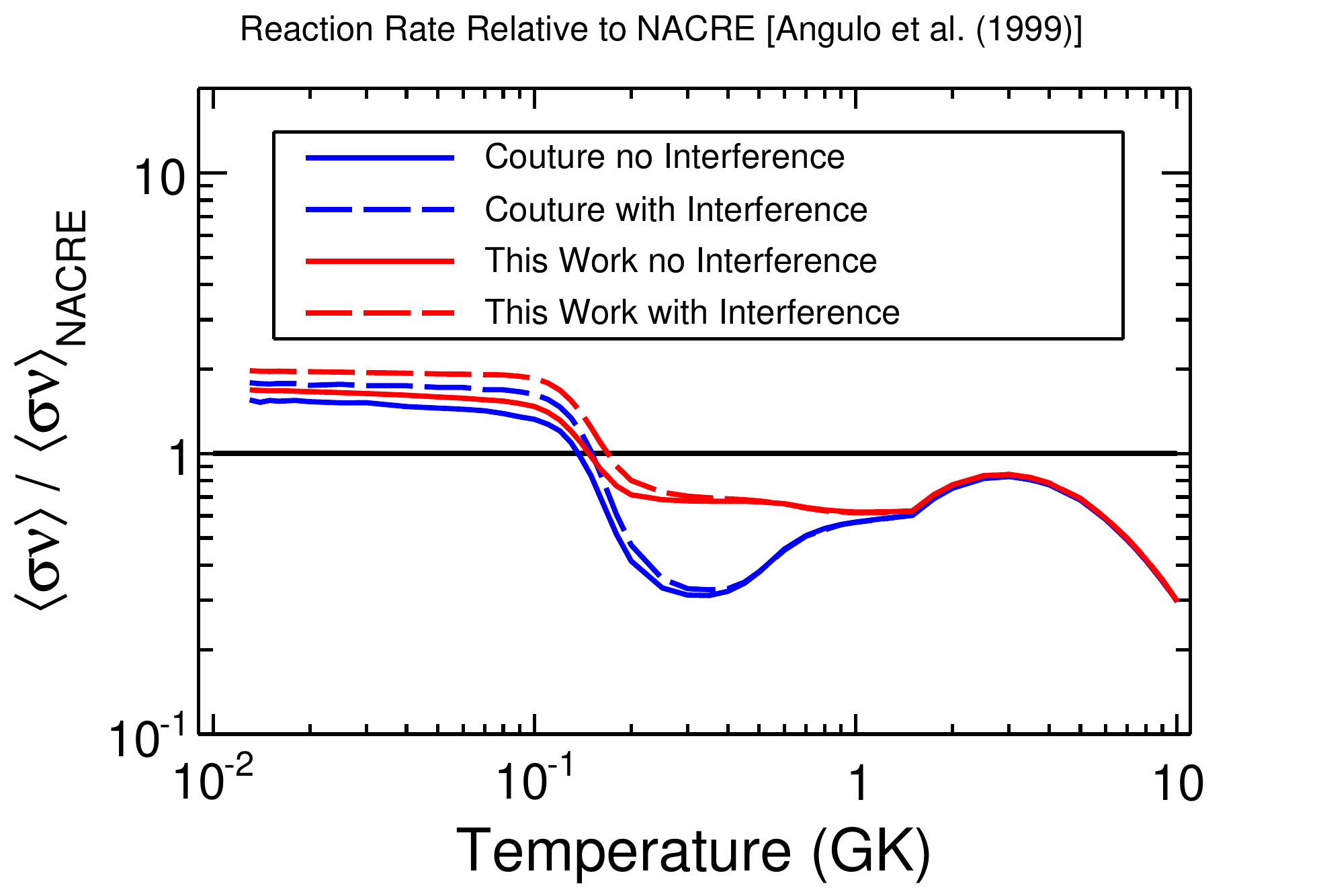}
\caption{$^{19}$F$(p,\gamma)^{20}$Ne reaction rate ratios relative to the NACRE compilation. The present work (red lines) are compared with that of Couture \textit{et al.} \cite{couture2008}. Rates were calculated both with interference terms included (dashed lines) and without (solid lines).} \label{fig:rates}
 \end{figure}

The $^{19}$F$(p,\gamma)^{20}$Ne reaction rate was calculated using the \texttt{RatesMC} code \cite{iliadis2015}. Strengths for resonances above the 634-keV resonance were taken from Angulo \textit{et al.} \cite{ANGULO1999}. Parameters for the 634-keV resonance and below, except the 323-keV resonance studied in this work, were adopted from Couture \textit{et al.} \cite{couture2008}. Interference effects were taken into account as prescribed by Couture \textit{et al.} Note that the \texttt{RatesMC} code as described in Ref. \cite{iliadis2015} only calculates interference terms for a pair of resonances and, therefore, had to be modified by the author of this work in order to calculate interference effects involving more than two resonances (in this case arising from three $1^{+}$ states). The non-resonant S-factor was adopted from Wiescher \textit{et al.} \cite{Wiescher1999} as $S(0) = 5.6$ keV.b, which the authors obtained by using a direct capture model combined with $^{20}$Ne bound-state spectroscopic factors from Ref. \cite{ajzenberg1987}. As displayed on Figure \ref{fig:rates}, our resulting rate is almost a factor of two larger than Couture \textit{et al.}, but lower than Angulo \textit{et al.} at 0.3 GK. The tabulated rate from this work is given in Table \ref{table:reactionrate} of Appendix \ref{sec:reactionrate}. The enhancement at low temperatures seen in both the present rate and that of Couture \textit{et al.} with respect to the NACRE rate is down to a larger assumed astrophysical S-factor for direct capture, which dominates the rate at temperatures below 0.1 GK. There is also clearly a some enhancement due to interference effects.

%\section{\label{sec:results}Astrophysical Impact}

%not sure if these sections are needed.

%\subsection{\label{sec:nova_calcs}Classical Novae}

% no impact

%\subsection{\label{sec:agb_stars}AGB Stars}

% no impact

\section{Conclusions}

In summary, we measured the strength of the 323-keV resonance in the $^{19}$F$(p,\gamma)^{20}$Ne reaction using inverse kinematics techniques. Owing to the ability to identify the reaction channel of interest via the $^{20}$Ne recoils we are able to effectively suppress copious $\gamma$-ray background from the far stronger $^{19}$F$(p,\alpha)^{16}$O channel. This enabled identification of both decay channels proceeding via the first $2^{-}$ state at 4966 keV as well as evidence of a strong direct to ground state transition, neither of which have been observed in any prior measurements of this reaction. Our final strength value of $\omega\gamma = 3.0^{+1.1}_{-0.9}$ meV is in good agreement with the literature, except the results of Subotic \textit{et al.} \cite{subotic1979}, and provided additional transitions observed here are added to the results of Couture \textit{et al.} \cite{couture2008}.

We calculate a new thermonuclear reaction rate adopting the present result for the 323-keV resonance and adopting literature values for all other resonances, as well as the interference effects studied by Couture \textit{et al.} \cite{couture2008}. Our new rate is intermediate between those of Couture \textit{et al.} and Angulo \textit{et al.} \cite{ANGULO1999}. Relative to the compilation by Angulo \textit{et al.}, the ratio of the $(p,\alpha)/(p,\gamma)$ reaction rates is decreased by a factor of two at $T<0.1$ GK to $2\times10^{3}$. The ratio climbs to $7\times10^{3}$ at $T\approx0.2$ GK, which is almost a factor of two lower than the ratio calculated at the same temperature with the Couture \textit{et al.} $^{19}$F$(p,\gamma)^{20}$Ne rate.

\section{Acknowledgements}

The authors thank the ISAC operations and technical staff at TRIUMF. TRIUMF’s core operations are supported via a contribution from the federal government through the National Research Council Canada, and the Government of British Columbia provides building capital funds. DRAGON is supported by funds from the Canadian Natural Science and Engineering Research Council (NSERC) under project number SAPPJ-2019-00039. MW gratefully acknowledges support from the Science and Technology Facilities Council (STFC). PA thanks the Claude Leon Foundation for support in the form of a Postdoctoral Fellowship. Authors from the Colorado School of Mines acknowledge funding via the U.S. Department of Energy grant DE\hyph{}FG02\hyph{}93ER40789. This article benefited from discussions within the ‘ChETEC’ COST Action (CA16117). The authors thank R. Longland of North Carolina State University for discussions relating to calculations of thermonuclear reaction rates presented in this work.

%J. Jos\'e acknowledges support from the Spanish MINECO grant AYA2017\hyph{}86274\hyph{}P, the EU FEDER funds and the AGAUR/ Generalitat de Catalunya grant SGR-661/2017.
%U. Battino acknowledges support from the European Research Council (ERC\hyph{}2015\hyph{}STG Nr.677497).

\bibliography{F19pg_bibliography.bib}

\appendix

\section{Thermonuclear Reaction Rate}
\label{sec:reactionrate}

This appendix contains the total thermonuclear reaction rate adopted following this work. The thermonuclear rate was computed using the \texttt{RatesMC} code, which calculates the log-normal parameters $\mu$ and $\sigma$ describing the reaction rate at a given temperature. Lower and upper rates are calculated at the 68\% confidence interval. The column labelled `A-D statistic' refers to the Anderson-Darling statistic, indicating how well a log-normal distribution describes the rate at a given temperature. An A-D statistic of less than $\approx1$ indicates that the rate is well described by a log-normal distribution. However, it has been shown that the assumption of a log-normal distributed reaction rate holds for A-D statistics in the $\approx 1-30$ range \cite{Iliadis-NPA841-2010}.

{\setlength{\tabcolsep}{8pt}

\begin{longtable*}{c c c c c c c}

\caption{Tabulated $^{19}$F$(p,\gamma)
^{20}$Ne total thermonuclear reaction rate determined from the present work, expressed in units of cm$^{3}$ mol$^{-1}$ s$^{-1}$.} 
% \noalign{\smallskip}

\label{table:reactionrate} \\

\toprule[1pt]\midrule[0.3pt]

 T [GK] & Low rate & Medium rate & High rate & Log-normal $\mu$ & Log-normal $\sigma$ & A-D statistic \\ \hline \noalign{\smallskip}
\endfirsthead

\multicolumn{7}{c}%
{{\tablename\ \thetable{} -- \textit{Continued from previous page}}} \\
\toprule[1pt]\midrule[0.3pt]
T [GK] & Low rate & Medium rate & High rate & Log-normal $\mu$ & Log-normal $\sigma$ & A-D statistic \\ \hline \noalign{\smallskip}
\endhead

\hline \noalign{\smallskip} \multicolumn{7}{c}{{\textit{Continued on next page}}} \\ \noalign{\smallskip} \hline
\endfoot

\midrule[0.3pt]\bottomrule[1pt]
\endlastfoot

0.010 & 6.60$\times$10$^{-28}$ & 8.17$\times$10$^{-28}$ &
      1.02$\times$10$^{-27}$ &  -6.237$\times$10$^{+01}$ &
       2.17$\times$10$^{-01}$  & 1.75$\times$10$^{+00}$ \\ 
0.011 & 8.66$\times$10$^{-27}$ & 1.07$\times$10$^{-26}$ &
      1.33$\times$10$^{-26}$ &  -5.980$\times$10$^{+01}$ &
       2.17$\times$10$^{-01}$  & 9.77$\times$10$^{-01}$ \\ 
0.012 & 8.40$\times$10$^{-26}$ & 1.03$\times$10$^{-25}$ &
      1.30$\times$10$^{-25}$ &  -5.752$\times$10$^{+01}$ &
       2.19$\times$10$^{-01}$  & 2.66$\times$10$^{+00}$ \\ 
0.013 & 6.42$\times$10$^{-25}$ & 7.89$\times$10$^{-25}$ &
      9.77$\times$10$^{-25}$ &  -5.549$\times$10$^{+01}$ &
       2.14$\times$10$^{-01}$  & 1.60$\times$10$^{+00}$ \\ 
0.014 & 3.99$\times$10$^{-24}$ & 4.93$\times$10$^{-24}$ &
      6.13$\times$10$^{-24}$ &  -5.366$\times$10$^{+01}$ &
       2.14$\times$10$^{-01}$  & 1.70$\times$10$^{+00}$ \\ 
0.015 & 2.11$\times$10$^{-23}$ & 2.61$\times$10$^{-23}$ &
      3.24$\times$10$^{-23}$ &  -5.200$\times$10$^{+01}$ &
       2.16$\times$10$^{-01}$  & 1.32$\times$10$^{+00}$ \\ 
0.016 & 9.64$\times$10$^{-23}$ & 1.20$\times$10$^{-22}$ &
      1.50$\times$10$^{-22}$ &  -5.047$\times$10$^{+01}$ &
       2.20$\times$10$^{-01}$  & 7.26$\times$10$^{-01}$ \\ 
0.018 & 1.43$\times$10$^{-21}$ & 1.76$\times$10$^{-21}$ &
      2.19$\times$10$^{-21}$ &  -4.778$\times$10$^{+01}$ &
       2.13$\times$10$^{-01}$  & 1.72$\times$10$^{+00}$ \\ 
0.020 & 1.45$\times$10$^{-20}$ & 1.79$\times$10$^{-20}$ &
      2.22$\times$10$^{-20}$ &  -4.547$\times$10$^{+01}$ &
       2.14$\times$10$^{-01}$  & 1.19$\times$10$^{+00}$ \\ 
0.025 & 1.52$\times$10$^{-18}$ & 1.87$\times$10$^{-18}$ &
      2.33$\times$10$^{-18}$ &  -4.082$\times$10$^{+01}$ &
       2.16$\times$10$^{-01}$  & 1.28$\times$10$^{+00}$ \\ 
0.030 & 5.24$\times$10$^{-17}$ & 6.45$\times$10$^{-17}$ &
      8.01$\times$10$^{-17}$ &  -3.727$\times$10$^{+01}$ &
       2.12$\times$10$^{-01}$  & 1.86$\times$10$^{+00}$ \\ 
0.040 & 9.11$\times$10$^{-15}$ & 1.11$\times$10$^{-14}$ &
      1.37$\times$10$^{-14}$ &  -3.212$\times$10$^{+01}$ &
       2.07$\times$10$^{-01}$  & 6.63$\times$10$^{-01}$ \\ 
0.050 & 3.56$\times$10$^{-13}$ & 4.38$\times$10$^{-13}$ &
      5.39$\times$10$^{-13}$ &  -2.846$\times$10$^{+01}$ &
       2.06$\times$10$^{-01}$  & 4.34$\times$10$^{-01}$ \\ 
0.060 & 5.92$\times$10$^{-12}$ & 7.19$\times$10$^{-12}$ &
      8.83$\times$10$^{-12}$ &  -2.565$\times$10$^{+01}$ &
       2.03$\times$10$^{-01}$  & 1.37$\times$10$^{+00}$ \\ 
0.070 & 5.49$\times$10$^{-11}$ & 6.65$\times$10$^{-11}$ &
      8.13$\times$10$^{-11}$ &  -2.343$\times$10$^{+01}$ &
       2.01$\times$10$^{-01}$  & 2.05$\times$10$^{+00}$ \\ 
0.080 & 3.48$\times$10$^{-10}$ & 4.21$\times$10$^{-10}$ &
      5.17$\times$10$^{-10}$ &  -2.158$\times$10$^{+01}$ &
       2.01$\times$10$^{-01}$  & 1.90$\times$10$^{+00}$ \\ 
0.090 & 1.67$\times$10$^{-09}$ & 2.02$\times$10$^{-09}$ &
      2.45$\times$10$^{-09}$ &  -2.002$\times$10$^{+01}$ &
       1.94$\times$10$^{-01}$  & 1.21$\times$10$^{+00}$ \\ 
0.100 & 6.49$\times$10$^{-09}$ & 7.76$\times$10$^{-09}$ &
      9.46$\times$10$^{-09}$ &  -1.867$\times$10$^{+01}$ &
       1.90$\times$10$^{-01}$  & 1.47$\times$10$^{+00}$ \\ 
0.110 & 2.14$\times$10$^{-08}$ & 2.57$\times$10$^{-08}$ &
      3.11$\times$10$^{-08}$ &  -1.747$\times$10$^{+01}$ &
       1.90$\times$10$^{-01}$  & 1.02$\times$10$^{+00}$ \\ 
0.120 & 6.21$\times$10$^{-08}$ & 7.49$\times$10$^{-08}$ &
      9.02$\times$10$^{-08}$ &  -1.641$\times$10$^{+01}$ &
       1.89$\times$10$^{-01}$  & 9.15$\times$10$^{-01}$ \\ 
0.130 & 1.64$\times$10$^{-07}$ & 1.98$\times$10$^{-07}$ &
      2.41$\times$10$^{-07}$ &  -1.543$\times$10$^{+01}$ &
       1.95$\times$10$^{-01}$  & 2.16$\times$10$^{+00}$ \\ 
0.140 & 4.14$\times$10$^{-07}$ & 4.97$\times$10$^{-07}$ &
      6.00$\times$10$^{-07}$ &  -1.451$\times$10$^{+01}$ &
       1.95$\times$10$^{-01}$  & 5.67$\times$10$^{+00}$ \\ 
0.150 & 9.85$\times$10$^{-07}$ & 1.18$\times$10$^{-06}$ &
      1.44$\times$10$^{-06}$ &  -1.364$\times$10$^{+01}$ &
       2.03$\times$10$^{-01}$  & 9.02$\times$10$^{+00}$ \\ 
0.160 & 2.29$\times$10$^{-06}$ & 2.77$\times$10$^{-06}$ &
      3.41$\times$10$^{-06}$ &  -1.278$\times$10$^{+01}$ &
       2.11$\times$10$^{-01}$  & 9.30$\times$10$^{+00}$ \\ 
0.180 & 1.16$\times$10$^{-05}$ & 1.46$\times$10$^{-05}$ &
      1.86$\times$10$^{-05}$ &  -1.113$\times$10$^{+01}$ &
       2.39$\times$10$^{-01}$  & 5.18$\times$10$^{+00}$ \\ 
0.200 & 5.36$\times$10$^{-05}$ & 6.92$\times$10$^{-05}$ &
      9.07$\times$10$^{-05}$ &  -9.571$\times$10$^{+00}$ &
       2.69$\times$10$^{-01}$  & 1.40$\times$10$^{+00}$ \\ 
0.250 & 1.13$\times$10$^{-03}$ & 1.53$\times$10$^{-03}$ &
      2.11$\times$10$^{-03}$ &  -6.472$\times$10$^{+00}$ &
       3.17$\times$10$^{-01}$  & 9.88$\times$10$^{-01}$ \\ 
0.300 & 9.44$\times$10$^{-03}$ & 1.33$\times$10$^{-02}$ &
      1.85$\times$10$^{-02}$ &  -4.324$\times$10$^{+00}$ &
       3.37$\times$10$^{-01}$  & 6.03$\times$10$^{-01}$ \\ 
0.350 & 4.44$\times$10$^{-02}$ & 6.16$\times$10$^{-02}$ &
      8.68$\times$10$^{-02}$ &  -2.779$\times$10$^{+00}$ &
       3.40$\times$10$^{-01}$  & 1.04$\times$10$^{+00}$ \\ 
0.400 & 1.42$\times$10$^{-01}$ & 1.98$\times$10$^{-01}$ &
      2.76$\times$10$^{-01}$ &  -1.618$\times$10$^{+00}$ &
       3.31$\times$10$^{-01}$  & 5.32$\times$10$^{-01}$ \\ 
0.450 & 3.63$\times$10$^{-01}$ & 4.93$\times$10$^{-01}$ &
      6.77$\times$10$^{-01}$ &  -7.002$\times$10$^{-01}$ &
       3.15$\times$10$^{-01}$  & 1.60$\times$10$^{+00}$ \\ 
0.500 & 8.22$\times$10$^{-01}$ & 1.08$\times$10$^{+00}$ &
      1.45$\times$10$^{+00}$ &  8.533$\times$10$^{-02}$ &
       2.80$\times$10$^{-01}$  & 2.13$\times$10$^{+00}$ \\ 
0.600 & 3.27$\times$10$^{+00}$ & 4.04$\times$10$^{+00}$ &
      5.07$\times$10$^{+00}$ &  1.404$\times$10$^{+00}$ &
       2.21$\times$10$^{-01}$  & 1.57$\times$10$^{+00}$ \\ 
0.700 & 1.03$\times$10$^{+01}$ & 1.22$\times$10$^{+01}$ &
      1.46$\times$10$^{+01}$ &  2.507$\times$10$^{+00}$ &
       1.79$\times$10$^{-01}$  & 8.90$\times$10$^{-01}$ \\ 
0.800 & 2.55$\times$10$^{+01}$ & 3.03$\times$10$^{+01}$ &
      3.60$\times$10$^{+01}$ &  3.411$\times$10$^{+00}$ &
       1.71$\times$10$^{-01}$  & 2.96$\times$10$^{-01}$ \\ 
0.900 & 5.37$\times$10$^{+01}$ & 6.36$\times$10$^{+01}$ &
      7.55$\times$10$^{+01}$ &  4.155$\times$10$^{+00}$ &
       1.71$\times$10$^{-01}$  & 4.38$\times$10$^{-01}$ \\ 
1.000 & 9.84$\times$10$^{+01}$ & 1.17$\times$10$^{+02}$ &
      1.39$\times$10$^{+02}$ &  4.762$\times$10$^{+00}$ &
       1.76$\times$10$^{-01}$  & 3.72$\times$10$^{-01}$ \\ 
1.250 & 2.97$\times$10$^{+02}$ & 3.53$\times$10$^{+02}$ &
      4.23$\times$10$^{+02}$ &  5.869$\times$10$^{+00}$ &
       1.79$\times$10$^{-01}$  & 9.49$\times$10$^{-01}$ \\ 
1.500 & 6.16$\times$10$^{+02}$ & 7.31$\times$10$^{+02}$ &
      8.70$\times$10$^{+02}$ &  6.596$\times$10$^{+00}$ &
       1.74$\times$10$^{-01}$  & 5.39$\times$10$^{-01}$ \\ 
1.750 & 1.03$\times$10$^{+03}$ & 1.21$\times$10$^{+03}$ &
      1.43$\times$10$^{+03}$ &  7.103$\times$10$^{+00}$ &
       1.64$\times$10$^{-01}$  & 8.37$\times$10$^{-01}$ \\ 
2.000 & 1.51$\times$10$^{+03}$ & 1.76$\times$10$^{+03}$ &
      2.08$\times$10$^{+03}$ &  7.480$\times$10$^{+00}$ &
       1.59$\times$10$^{-01}$  & 8.50$\times$10$^{-01}$ \\ 
2.500 & 2.56$\times$10$^{+03}$ & 2.95$\times$10$^{+03}$ &
      3.43$\times$10$^{+03}$ &  7.995$\times$10$^{+00}$ &
       1.46$\times$10$^{-01}$  & 2.17$\times$10$^{+00}$ \\ 
3.000 & 3.63$\times$10$^{+03}$ & 4.12$\times$10$^{+03}$ &
      4.71$\times$10$^{+03}$ &  8.329$\times$10$^{+00}$ &
       1.31$\times$10$^{-01}$  & 1.80$\times$10$^{+00}$ \\ 
3.500 & 4.56$\times$10$^{+03}$ & 5.15$\times$10$^{+03}$ &
      5.84$\times$10$^{+03}$ &  8.549$\times$10$^{+00}$ &
       1.25$\times$10$^{-01}$  & 1.45$\times$10$^{+00}$ \\ 
4.000 & 5.38$\times$10$^{+03}$ & 6.04$\times$10$^{+03}$ &
      6.80$\times$10$^{+03}$ &  8.709$\times$10$^{+00}$ &
       1.18$\times$10$^{-01}$  & 1.38$\times$10$^{+00}$ \\ 
5.000 & 6.55$\times$10$^{+03}$ & 7.28$\times$10$^{+03}$ &
      8.11$\times$10$^{+03}$ &  8.894$\times$10$^{+00}$ &
       1.07$\times$10$^{-01}$  & 5.64$\times$10$^{-01}$ \\ 
6.000 & 7.20$\times$10$^{+03}$ & 7.94$\times$10$^{+03}$ &
      8.77$\times$10$^{+03}$ &  8.981$\times$10$^{+00}$ &
       1.00$\times$10$^{-01}$  & 5.30$\times$10$^{-01}$ \\ 
7.000 & 7.41$\times$10$^{+03}$ & 8.21$\times$10$^{+03}$ &
      9.09$\times$10$^{+03}$ &  9.014$\times$10$^{+00}$ &
       1.00$\times$10$^{-01}$  & 6.30$\times$10$^{-01}$ \\ 
8.000 & 7.44$\times$10$^{+03}$ & 8.22$\times$10$^{+03}$ &
      9.07$\times$10$^{+03}$ &  9.014$\times$10$^{+00}$ &
       1.01$\times$10$^{-01}$  & 1.03$\times$10$^{+00}$ \\ 
9.000 & 7.33$\times$10$^{+03}$ & 8.06$\times$10$^{+03}$ &
      8.92$\times$10$^{+03}$ &  8.998$\times$10$^{+00}$ &
       9.99$\times$10$^{-02}$  & 1.42$\times$10$^{+00}$ \\ 
10.000 & 7.12$\times$10$^{+03}$ & 7.86$\times$10$^{+03}$ &
      8.68$\times$10$^{+03}$ &  8.971$\times$10$^{+00}$ &
       9.93$\times$10$^{-02}$  & 1.15$\times$10$^{+00}$ \\ 

\end{longtable*}
}

\end{document}